\documentclass[10pt]{article}
\usepackage{graphicx}
\usepackage{amsmath}
\usepackage{amssymb}
\usepackage{caption2}
\begin{document}
\bibliographystyle{prsty}
\begin{center}
{\large {\bf \sc{  Next-to-leading order perturbative contributions  in the QCD sum rules for mesonic two-point correlation functions with unequal quark
masses  }}} \\[2mm]
Zhi-Gang Wang \footnote{E-mail,zgwang@aliyun.com.  }     \\
 Department of Physics, North China Electric Power University,
Baoding 071003, P. R. China
\end{center}

\begin{abstract}
In this article, we assume that the two quarks have unequal masses and calculate the next-to-leading order contributions  to the spectral densities of the mesonic two-point correlation functions of the vector, axialvector, scalar and pseudoscalar currents. We take dimensional regularization to regularize both the ultraviolet and infrared divergences, and use    optical theorem to obtain the spectral densities directly, furthermore, we present some necessary technical details  for readers convenience.    The analytical expressions are applicable in many phenomenological analysis besides  the QCD sum rules.
\end{abstract}



\section{Introduction}

The QCD sum rules is a powerful theoretical tool  in
 studying   the ground state mesons \cite{SVZ79,Reinders85}, and has given many successful descriptions of the meson properties. In the case of the interpolating currents consist of quarks with equal masses or zero masses,  the leading order and next-to-leading order perturbative contributions to the spectral densities of the mesonic two-point correlation functions are  known \cite{Reinders85,NarisonBook}.  There are few works on the next-to-leading order perturbative contributions to the
    spectral densities of the interpolating currents for unequal quark masses,  as the calculations are very difficult.

   In Refs.\cite{Reinders1980,Reinders1981}, Reinders, Rubinstein and Yazaki calculate the next-to-leading order  contributions to the spectral densities
of the mesonic two-point correlation functions for the pseudoscalar and vector currents in the unequal mass case.  Some technical details are presented in Ref.\cite{Reinders85}.
In the  famous review (see Ref.\cite{Reinders85}), Reinders, Rubinstein and Yazaki illustrate how to use optical theorem to calculate the next-to-leading order contributions to spectral densities of the mesonic two-point correlation functions (for the scalar, pseudoscalar, vector and axialvector currents) directly, and regularize the  divergences in  the massive-gluon scheme.  In Ref.\cite{Schilcher81}, Schilcher,  Tran and Nasrallah also resort to the massive-gluon scheme to regularize the infrared divergence, and use optical theorem to   calculate the next-to-leading order contributions to the longitudinal and transverse spectral densities of the mesonic two-point correlation functions directly for arbitrary conserved or non-conserved vector or   axialvector currents of the massive quarks, and obtain analytical expressions. Although those works are all performed in the massive-gluon scheme, the resulting (infrared) divergent terms  are different from each other, the expressions also different from each other.

In Ref.\cite{Generalis1990}, Generalis calculates the two-loop Feynman diagrams directly  using the dimensional regularization, and obtain the next-to-leading order contributions to the  mesonic two-point correlation functions for the vector and axialvector currents with unequal quark masses.
     We have to obtain the spectral densities through dispersion relation from the cumbersome expressions, it is a difficult work.

In this article, we resort to optical theorem (or Cutkosky's rule) to calculate the next-to-leading order contributions to the spectral densities of the mesonic two-point correlation functions for the scalar, pseudoscalar, vector and axialvector currents in the case of unequal quark masses. In calculations, we regularize both the ultraviolet and infrared divergences  in the scheme of dimensional regularization, and present the necessary technical details for the readers convenience, i.e. one can check the calculations. There are two routines in application of  optical theorem, we resort to the routine used in Refs.\cite{Reinders85,Schilcher81}, not the one used in Ref.\cite{Cut-2}.  After the present work is finished and submitted to the net http://arxiv.org/, some friends kindly draw my attention to Refs.\cite{Djouadi1994,Groote1301}. In  Ref.\cite{Djouadi1994}, Djouadi and Gambino  study the QCD corrections to the  electro-weak gauge boson self-energies and Higgs self-energies for arbitrary momentum transfer and for different internal quark masses, and obtain explicit
 expressions  for both the real and imaginary parts of the self-energies. They calculate the two-loop Feynman diagrams directly  using the dimensional regularization, the tedious work is difficult to follow. On the other hand, if we calculate the imaginary parts of  the mesonic two-point correlation functions (or the self-energies) directly via optical theorem, the calculations are greatly simplified and more easy to follow,  furthermore, we can separate the contributions come from the real and  virtual gluon emissions explicitly,  which have direct applications in studying the decays of a polarized $W$ boson into massive quark-antiquark pairs, such as $t \to b W^+ \to b q_1\bar{q}_2$, $H \to W^-W^{*+} \to W^- q_1\bar{q}_2$, the $H$ denotes the Higgs boson \cite{Groote1301}.

The article is arranged as follows:  we calculate the next-to-leading order contributions to spectral densities of  the mesonic two-point correlation functions in Sect.2;
in Sect.3, we present the final analytical expressions; and Sect.4 is reserved for our
conclusions.

\section{Explicit calculations of the spectral densities at the next-to-leading order}
In the following, we write down  the mesonic two-point correlation functions in the QCD sum rules,
\begin{eqnarray}
\Pi_{\mu\nu}(p)&=&i\int d^4x e^{ip \cdot x} \langle 0|T\left\{J_{\mu}(x)J_{\nu}^{\dagger}(0)\right\}|0\rangle \, , \nonumber\\
&=&\Pi_{V/A}(p)\left(-g_{\mu\nu}+\frac{p^\mu p^\nu}{p^2} \right) +\Pi_{S/P}(p)\frac{p^\mu p^\nu}{p^2} \, ,\nonumber\\
\Pi_{S/P}(p)&=&\frac{(m_1\mp m_2)^2}{p^2}\widetilde{\Pi}_{S/P}(p)\, ,\nonumber\\
\widetilde{\Pi}_{S/P}(p)&=&i\int d^4x e^{ip \cdot x} \langle 0|T\left\{J(x)J^{\dagger}(0)\right\}|0\rangle \, ,
\end{eqnarray}
where $J^{\mu}(x)=J_V^{\mu}(x),J_A^{\mu}(x)$ and $J(x)=J_S(x),J_P(x)$. The lower indexes
 denote the scalar ($S$), pseudoscalar ($P$), vector ($V$) and axialvector ($A$) currents respectively. The correlation functions can
 be expressed in the following form through dispersion relation,
 \begin{eqnarray}
\Pi_i(p^2)&=&\frac{1}{\pi}\int_{\Delta^2}^\infty ds \frac{{\rm Im\Pi}_i(s)}{s-p^2}\, ,
\end{eqnarray}
where $i=S,P,V,A$, the $\Delta^2=(m_1+m_2)^2$ is the threshold parameter, and
\begin{eqnarray}
\frac{{\rm Im}\Pi_{i}(s)}{\pi}&=&\rho^0_{i}(s)+\rho^1_{i}(s)+\rho^2_{i}(s)+\cdots\, ,
\end{eqnarray}
the $\rho^0_{i}(s)$, $\rho^1_{i}(s)$, $\rho^2_{i}(s)$, $\cdots$ are the spectral densities of the leading order, next-to-leading order, and next-to-next-to-leading order, $\cdots$. At the leading order, the perturbative spectral densities are rather simple,
\begin{eqnarray}
 \rho^0_{V/A}(s)&=&\frac{3}{8\pi^2}\frac{\sqrt{\lambda(s,m_1^2,m_2^2)}}{s}\left\{s-(m_1\mp m_2)^2-\frac{\lambda(s,m_1^2,m_2^2)}{3s} \right\}\, , \nonumber\\
 \rho^0_{S/P}(s)&=&\frac{(m_1\mp m_2)^2}{s}\frac{3}{8\pi^2}\frac{\sqrt{\lambda(s,m_1^2,m_2^2)}}{s}\left\{s-(m_1\pm m_2)^2 \right\}\,, \end{eqnarray}
where
\begin{eqnarray}
\lambda(s,m_1^2,m_2^2)&=&s^2+m_1^4+m_2^4-2sm_1^2-2sm_2^2-2m_1^2m_2^2\, .
\end{eqnarray}

At the next-to-leading order, there are three Feynman diagrams contribute to the correlation functions, see Fig.1.
We calculate the imaginary parts (or the spectral densities) using  the Cutkosky's rule or optical theorem, the two approaches lead to the same results,  then use the dispersion relation to obtain the correlation functions. There are ten
possible cuts, six cuts  attribute to virtual gluon emissions, see Fig.2, and four cuts attribute to real gluon emissions, see Fig.3.

\begin{figure}
 \centering
 \includegraphics[totalheight=2.7cm,width=14cm]{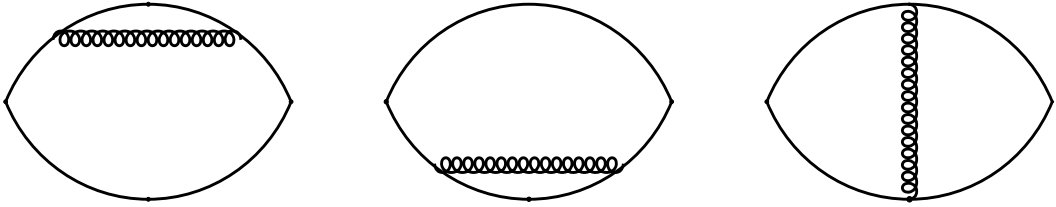}
    \caption{The next-to-leading order contributions to the correlation functions. }
\end{figure}

\begin{figure}
 \centering
 \includegraphics[totalheight=3.7cm,width=14cm]{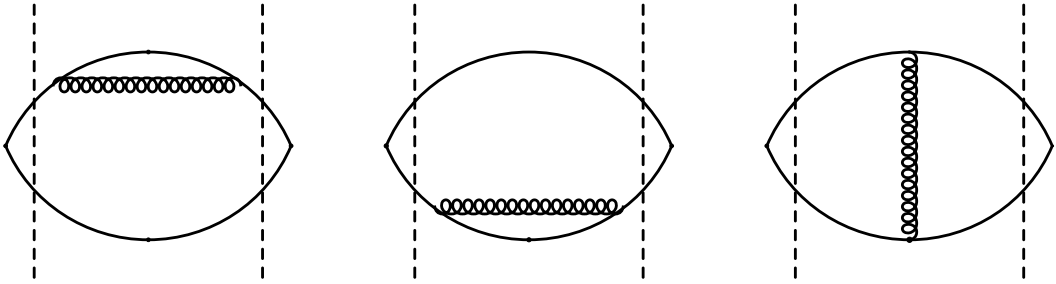}
    \caption{Six possible cuts correspond to virtual gluon emissions. }
\end{figure}

\begin{figure}
 \centering
 \includegraphics[totalheight=3.7cm,width=14cm]{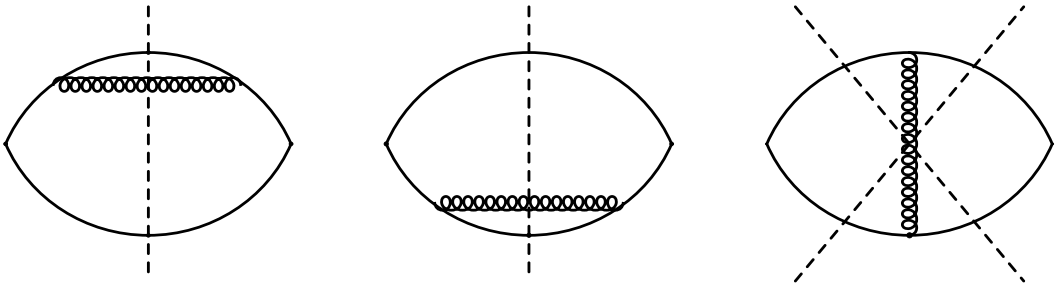}
    \caption{Four possible cuts correspond to  real gluon emissions. }
\end{figure}

\subsection{ Contributions of the virtual gluon emissions }
The six cuts shown in Fig.2  attribute to virtual gluon emissions and correspond to the self-energy corrections and vertex corrections, respectively.
The first four cuts corresponding to one-loop fermion's self-energy corrections, we calculate the Feynman diagrams directly  using the dimensional regularization and
choose the on-shell renormalization scheme to subtract the divergences so as to implement the wave-function renormalization and mass renormalization.
Then we can take into account all  contributions come from the six cuts shown in Fig.2 by the following simple replacement for each vertex in the interpolating currents\footnote{Here we use the vector current to
illustrate the procedure. Other currents can be treated analogously.},
\begin{eqnarray}
\bar{u}(p_1)\gamma_\mu u(p_2)&\rightarrow& \bar{u}(p_1)\gamma_\mu u(p_2)+\bar{u}(p_1)\widetilde{\Gamma}_\mu u(p_2)   \nonumber \\
&=& \sqrt{Z_1}\sqrt{Z_2}\bar{u}(p_1)\gamma_\mu u(p_2)+\bar{u}(p_1)\Gamma_\mu u(p_2)\nonumber \\
&=& \bar{u}(p_1)\gamma_\mu u(p_2)\left(1+\frac{1}{2}\delta Z_1+\frac{1}{2}\delta Z_2 \right)+\bar{u}(p_1)\Gamma_\mu u(p_2)\, ,
\end{eqnarray}
where
\begin{eqnarray}
Z_i&=&1+\delta Z_{i}=1+\frac{4}{3}\frac{\alpha_s}{\pi}\left(-\frac{1}{4\varepsilon_{\rm UV}}+\frac{1}{2\varepsilon_{\rm IR}}+\frac{3}{4}\log\frac{m_i^2}{4\pi\mu^2}+\frac{3}{4}\gamma-1\right)\, ,
\end{eqnarray}
is the $i$ quark's wave-function renormalization constant due to  the self-energy correction, see Fig.4,
and
\begin{eqnarray}
\Gamma_\mu &=&\gamma_\mu \frac{4}{3}g_s^2\int_0^1 dx \int_0^{1-x}dy \int \frac{d^D k_E}{(2\pi)^D} \nonumber\\
&& \frac{\Gamma(3)}{\left[k^2_E+(x p_1+y p_2)^2\right]^3}\left\{k^2_E\left(1-\frac{3}{2}\varepsilon_{\rm UV} \right)+2(1-x)(1-y)(s-m_1^2-m_2^2) \right.\nonumber\\
&&\left. +2(x+y)m_1m_2+2x(1-x)m_1^2+2y(1-y)m_2^2\right\}+\frac{4}{3}g_s^2\int_0^1 dx \int_0^{1-x}dy\int \frac{d^D k_E}{(2\pi)^D}\nonumber\\
&&\frac{4\Gamma(3)\left\{ \left[x^2m_1-y(1-x)m_2\right]p_{1\mu}+\left[y^2m_2-x(1-y)m_1\right]p_{2\mu}\right\}}{\left[k^2_E+(x p_1+y p_2)^2\right]^3}\, ,
\end{eqnarray}
is the vertex correction after performing the Wick rotation, see Fig.5. Here $\gamma$ is the Euler constant, $\mu^2$ is the renormalization  scale, and the Euclidean momentum $k_E=(k_1,k_2,k_3,k_4)$.  In this article, we take the dimension  $D=4-2\varepsilon_{\rm UV}=4+2\varepsilon_{\rm IR}$ to regularize the ultraviolet and infrared divergences respectively, and add the renormalization  scale factors $\mu^{ 2\varepsilon_{\rm UV}}$ or $\mu^{ -2\varepsilon_{\rm IR}}$ when necessary. In the limit $m_1=m_2=m$, the $\Gamma_\mu$ can be reduced as
\begin{eqnarray}
\Gamma_\mu &=&\gamma_\mu \frac{4}{3}g_s^2\int_0^1 dx \int_0^{1-x}dy \int \frac{d^D k_E}{(2\pi)^D} \nonumber\\
&& \frac{\Gamma(3)}{\left[k^2_E+(x p_1+y p_2)^2\right]^3}\left\{k^2_E\left(1-\frac{3}{2}\varepsilon_{\rm UV} \right)+2(1-(x+y)+xy)(s-2m^2) \right.\nonumber\\
&&\left. +4(x+y)m^2-2(x+y)^2m^2+4xym^2\right\}+\frac{4}{3}g_s^2\int_0^1 dx \int_0^{1-x}dy\int\frac{d^D k_E}{(2\pi)^D}\nonumber\\
&&\frac{2\Gamma(3)m \left[(x+y)^2-(x+y)\right](p_{1\mu}+p_{2\mu})}{\left[k^2_E+(x p_1+y p_2)^2\right]^3}\, ,
\end{eqnarray}
then the resulting expression is greatly simplified.

We carry out the integral over the variables $x$, $y$ and $k_E$, and observe that the  ultraviolet divergent terms $\frac{1}{\varepsilon_{\rm UV}}$ in the $\Gamma_\mu$, $\delta Z_1$ and $\delta Z_2$ are canceled out with each other, which is a
consequence of the Ward identity. The net contributions have no ultraviolet divergence,
\begin{eqnarray}
\widetilde{\Gamma}_\mu &=& \frac{1}{3}\frac{\alpha_s}{\pi}\gamma_\mu f(s)+\frac{1}{3}\frac{\alpha_s}{\pi}f_{1}(s)p_{1\mu}+\frac{1}{3}\frac{\alpha_s}{\pi}f_{2}(s)p_{2\mu} \, ,
\end{eqnarray}
where
\begin{eqnarray}
f(s) &=& \overline{f}(s)+ \frac{2}{\varepsilon_{\rm IR}} +\log\frac{4\pi\mu^2}{s}+2\gamma-3+3\log\frac{m_1m_2}{4\pi\mu^2}-\frac{2(s-m_1^2-m_2^2)}{\sqrt{\lambda(s,m_1^2,m_2^2)}}\log\left(\frac{1+\omega}{1-\omega}\right)  \nonumber\\
&& \left(\frac{1}{\varepsilon_{\rm IR}} +\log\frac{s}{4\pi\mu^2}+\gamma\right)\, ,\nonumber\\
\overline{f}_{\pm}(s)&=& \overline{V}(s)+2(s-m_1^2-m_2^2)\left[\overline{V}_{00}(s)-V_{10}(s)-V_{01}(s)+V_{11}(s)\right] \pm 2m_1m_2\nonumber\\
&&\left[ V_{10}(s)+V_{01}(s)\right]+2m_1^2\left[ V_{10}(s)-V_{20}(s)\right]+2m_2^2\left[ V_{01}(s)-V_{02}(s)\right]\, ,\nonumber\\
f_{1\pm}(s)&=&4m_1V_{20}(s)\mp 4m_2V_{01}(s)\pm 4m_2V_{11}(s)\, ,\nonumber\\
f_{2\pm}(s)&=&\pm 4m_2V_{02}(s)-4m_1V_{10}(s)+4m_1V_{11}(s)\, , \nonumber\\
\omega&=&\sqrt{\frac{s-(m_1+m_2)^2}{s-(m_1-m_2)^2}}\, ,
\end{eqnarray}
and $s=p^2$, the $\overline{V}(s)$, $\overline{V}_{00}(s)$ and $V_{ij}(s)$ with $i,j=0,1,2$ are given explicitly in the appendix. Here we introduce the lower-indexes  $\pm$ to denote the corresponding expressions of the vector ($+$) and axialvector ($-$) currents respectively.
The infrared divergence  $\frac{1}{\varepsilon_{\rm IR}}$  comes from
 the wave-function renormalization constants $Z_i$ in Eq.(7),  while the infrared divergence  $\log\left(\frac{1+\omega}{1-\omega}\right)\frac{1}{\varepsilon_{\rm IR}}$ comes from  the term
\begin{eqnarray}
\gamma_\mu \frac{4}{3}g_s^2\int_0^1 dx \int_0^{1-x}dy \int \frac{d^D k_E}{(2\pi)^D}  \frac{2\Gamma(3)(s-m_1^2-m_2^2)}{\left[k^2_E+(x p_1+y p_2)^2\right]^3}
\end{eqnarray}
in Eq.(8). The infrared divergent terms obtained in the present work are consistent with that obtained by  Groote,  K\"orner and Tuvike in Ref.\cite{Groote1301}.
If we regularize the infrared divergences with a supposed gluon mass $m_g$ in the quantum field theory, the infrared divergent terms appear as $\log\frac{m_g^2}{\Lambda^2}$ in stead of $\frac{1}{\varepsilon_{\rm IR}}$, where the $\Lambda$ has a dimension of mass.
So in the massive-gluon scheme, we expect that the infrared   divergences   $\frac{1}{\varepsilon_{\rm IR}}$ and $\log\left(\frac{1+\omega}{1-\omega}\right)\frac{1}{\varepsilon_{\rm IR}}$ appear in the forms $\log\frac{m_g^2}{m_1m_2}$ and $\log\left(\frac{1+\omega}{1-\omega}\right)\log\frac{m_g^2}{m_1m_2}$, respectively.
 In Ref.\cite{Reinders85},  there only exists the infrared divergence $\log\frac{m_g^2}{m_1m_2}$; while in Ref.\cite{Schilcher81},  there exist the infrared divergences $\log\left(\frac{1+\omega}{1-\omega}\right)\log\frac{m_g^2}{m_1m_2}$ and $\log\frac{m_1}{m_2}\log\frac{m_g^2}{m_1m_2}$. In this article, we calculate the vertex corrections using the conventional Feynman parameters, while in Ref.\cite{Schilcher81}, Schilcher,  Tran and Nasrallah  calculate the vertex corrections using optical theorem and dispersion relation.

\begin{figure}
 \centering
 \includegraphics[totalheight=2cm,width=8cm]{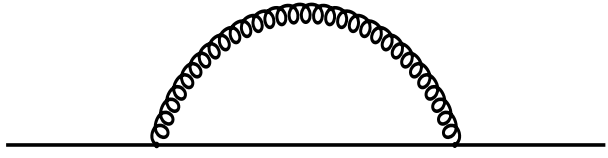}
    \caption{The quark self-energy correction. }
\end{figure}

\begin{figure}
 \centering
 \includegraphics[totalheight=3cm,width=4cm]{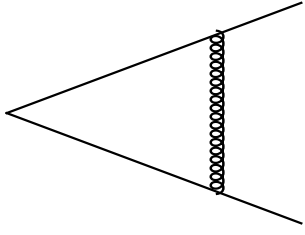}
    \caption{The vertex correction. }
\end{figure}

The total contributions of  the virtual gluon emissions (see Fig.2) to imaginary parts of the correlation functions can be expressed in the following form,
\begin{eqnarray}
\frac{{\rm Im}\Pi^V_{\mu\nu}(s)}{\pi}&=&\frac{4}{3}\frac{\alpha_s}{\pi}\frac{3}{\pi}\int \frac{d^{D-1}\vec{p}_1}{(2\pi)^{D-1}2E_{p_1}}\frac{d^{D-1}\vec{p}_2}{(2\pi)^{D-1}2E_{p_2}}(2\pi)^D \delta^D(p-p_1-p_2)\nonumber\\
&&\left\{ f(s)\left[p_{1\mu}p_{2\nu}+ p_{2\mu}p_{1\nu}-\frac{s-(m_1-m_2)^2}{2}g_{\mu\nu}\right]+f_1(s)(m_1p_{1\mu}p_{2\nu}-m_2p_{1\mu}p_{1\nu})\right.\nonumber\\
&&\left. +f_2(s)(m_2p_{2\mu}p_{1\nu}-m_1p_{2\mu}p_{2\nu})\right\} \, .
\end{eqnarray}
In this article, the up-indexes $V$ and $R$ denote the contributions come from the virtual and real gluon emissions respectively.
We carry out the integrals in Eq.(13) directly in $D=4+2\varepsilon_{\rm IR}$ dimension as there is no ultraviolet divergence, and obtain the following results,
\begin{eqnarray}
\frac{{\rm Im}\Pi^V_{V/A}(s)}{\pi}&=&\frac{4}{3}\frac{\alpha_s}{\pi}\rho_{V/A}^0(s)\left\{\frac{1}{\varepsilon_{\rm IR}}-2\log4\pi+2\gamma-\frac{7}{2}+\frac{1}{2}\log\frac{\lambda^2(s,m_1^2,m_2^2)m_1^3m_2^3}{\mu^8 s^3} +\frac{1}{2}\overline{f}_{\pm}(s)\right.\nonumber\\
&&\left.-\frac{s-m_1^2-m_2^2}{\sqrt{\lambda(s,m_1^2,m_2^2)}}\log\left(\frac{1+\omega}{1-\omega}\right)\left[\frac{1}{\varepsilon_{\rm IR}}-2\log4\pi+2\gamma-2+\log\frac{\lambda(s,m_1^2,m_2^2)}{\mu^4} \right] \right\} \nonumber\\
&&+\frac{4}{3}\frac{\alpha_s}{\pi}\frac{\sqrt{\lambda(s,m_1^2,m_2^2)}^3}{s^2}\left\{ \frac{1}{12\pi^2} \left[1-\frac{s-m_1^2-m_2^2}{\sqrt{\lambda(s,m_1^2,m_2^2)}}\log\left(\frac{1+\omega}{1-\omega}\right) \right]\right.\nonumber\\
&&\left.-\frac{(m_1\pm m_2)\left[f_{1\pm}(s)+f_{2\pm}(s) \right]}{32\pi^2} \right\} \, ,
\end{eqnarray}
\begin{eqnarray}
\frac{{\rm Im}\Pi^V_{S/P}(s)}{\pi}&=&\frac{4}{3}\frac{\alpha_s}{\pi}\rho_{S/P}^0(s)\left\{\frac{1}{\varepsilon_{\rm IR}}-2\log4\pi+2\gamma-\frac{7}{2}+\frac{1}{2}\log\frac{\lambda^2(s,m_1^2,m_2^2)m_1^3m_2^3}{\mu^8 s^3} +\frac{1}{2}\overline{f}_{\pm}(s)\right.\nonumber\\
&&\left.-\frac{s-m_1^2-m_2^2}{\sqrt{\lambda(s,m_1^2,m_2^2)}}\log\left(\frac{1+\omega}{1-\omega}\right)\left[\frac{1}{\varepsilon_{\rm IR}}-2\log4\pi+2\gamma-2+\log\frac{\lambda(s,m_1^2,m_2^2)}{\mu^4} \right] \right\} \nonumber\\
&&+\frac{4}{3}\frac{\alpha_s}{\pi}\frac{3}{32\pi^2}\frac{\sqrt{\lambda(s,m_1^2,m_2^2)}}{s^2}\left\{(s+m_1^2-m_2^2)(s+m_2^2-m_1^2)\left[f_{1\pm}(s)m_1\pm f_{2\pm}(s)m_2\right]  \right.\nonumber\\
&&\left.\mp f_{1\pm}(s)m_2(s+m_1^2-m_2^2)^2-f_{2\pm}(s)m_1(s+m_2^2-m_1^2)^2 \right\} \, ,
\end{eqnarray}
where we have used the $\gamma_5$ symmetry of the interpolating currents to obtain the  spectral densities $\frac{{\rm Im}\Pi^V_{A}(s)}{\pi}$ and $\frac{{\rm Im}\Pi^V_{P}(s)}{\pi}$.

\subsection{ Contributions of the real gluon emissions }
The four cuts shown in Fig.3 correspond to real gluon emissions. The scattering amplitudes for the real gluon emissions are shown explicitly in Fig.6. From Fig.6,
we can write down the scattering amplitude $T^a_{\mu\alpha}(p)$,
\begin{eqnarray}
T^a_{\mu\alpha}(p)&=&\bar{u}(p_1)\left\{ ig_s \frac{\lambda^a}{2}\gamma_\alpha \frac{i}{\!\not\!{p}_1+\!\not\!{k}-m_1}\gamma_\mu+\gamma_\mu\frac{i}{-\!\not\!{p}_2-\!\not\!{k}-m_2}ig_s\frac{\lambda^a}{2}\gamma_\alpha\right\}v(p_2)\,  ,
\end{eqnarray}
where $\lambda^a$ is the Gell-Mann matrix. Then we obtain the corresponding contributions ${\rm Im}\Pi^R_V(s)$ and ${\rm Im}\Pi^R_S(s)$ to the imaginary parts  of the correlation function with
 optical theorem,
\begin{eqnarray}
\frac{{\rm Im}\Pi^R_V(s)}{\pi}&=&-\frac{1}{2\pi(D-1)}  \int \frac{d^{D-1}\vec{k}}{(2\pi)^{D-1}2E_k}\frac{d^{D-1}\vec{p}_1}{(2\pi)^{D-1}2E_{p_1}}\frac{d^{D-1}\vec{p}_2}{(2\pi)^{D-1}2E_{p_2}}(2\pi)^{D}\delta^D(p-k-p_1-p_2)\nonumber\\
&&{\rm Tr}\left\{T^a_{\mu\alpha}(p)T^{a\dagger}_{\nu\beta}(p)\right\}g^{\alpha\beta}\left( -g^{\mu\nu}+\frac{p^\mu p^\nu}{p^2} \right)   \nonumber\\
&=&-\frac{2g_s^2}{\pi(D-1)}  \int \frac{d^{D-1}\vec{k}}{(2\pi)^{D-1}2E_k}\frac{d^{D-1}\vec{p}_1}{(2\pi)^{D-1}2E_{p_1}}\frac{d^{D-1}\vec{p}_2}{(2\pi)^{D-1}2E_{p_2}}(2\pi)^{D}\delta^D(p-k-p_1-p_2)\nonumber\\
&&\left\{6\left[\frac{m_1^2}{(k\cdot p_1)^2} +\frac{m_2^2}{(k\cdot p_2)^2}-\frac{s-m_1^2-m_2^2}{k\cdot p_1 k\cdot p_2}+\frac{s-K^2}{k\cdot p_1 k\cdot p_2}\right]\left[ s-(m_1-m_2)^2\right. \right.\nonumber\\
&&\left.-\frac{\lambda(s,m_1^2,m_2^2)}{3s}\right]+4\varepsilon_{\rm IR}\left[s-(m_1-m_2)^2\right]\left[\frac{m_1^2}{(k\cdot p_1)^2} +\frac{m_2^2}{(k\cdot p_2)^2}-\frac{s-m_1^2-m_2^2}{k\cdot p_1 k\cdot p_2} \right]\nonumber\\
&&\left.-\frac{(s-K^2)^2}{k\cdot p_1 k\cdot p_2}\left[2+\frac{(m_1-m_2)^2}{s}\right]+16\right\} \, ,
\end{eqnarray}
\begin{eqnarray}
\frac{{\rm Im}\Pi^R_S(s)}{\pi}&=&-\frac{1}{2\pi}  \int \frac{d^{D-1}\vec{k}}{(2\pi)^{D-1}2E_k}\frac{d^{D-1}\vec{p}_1}{(2\pi)^{D-1}2E_{p_1}}\frac{d^{D-1}\vec{p}_2}{(2\pi)^{D-1}2E_{p_2}}(2\pi)^{D}\delta^D(p-k-p_1-p_2)\nonumber\\
&&{\rm Tr}\left\{T^a_{\mu\alpha}(p)T^{a\dagger}_{\nu\beta}(p)\right\}g^{\alpha\beta}\frac{p^\mu p^\nu}{p^2}   \nonumber\\
&=&-\frac{2g_s^2}{\pi}  \int \frac{d^{D-1}\vec{k}}{(2\pi)^{D-1}2E_k}\frac{d^{D-1}\vec{p}_1}{(2\pi)^{D-1}2E_{p_1}}\frac{d^{D-1}\vec{p}_2}{(2\pi)^{D-1}2E_{p_2}}(2\pi)^{D}\delta^D(p-k-p_1-p_2)\nonumber\\
&&\frac{(m_1-m_2)^2}{s}\left\{2\left[ s-(m_1+m_2)^2\right]\left[\frac{m_1^2}{(k\cdot p_1)^2} +\frac{m_2^2}{(k\cdot p_2)^2}-\frac{s-m_1^2-m_2^2}{k\cdot p_1 k\cdot p_2}\right.\right.\nonumber\\
&&\left.\left.+\frac{s-K^2}{k\cdot p_1 k\cdot p_2}\right]  -\frac{(s-K^2)^2}{k\cdot p_1 k\cdot p_2}\right\} \, ,
\end{eqnarray}
where we have used the identities $\sum u(p_1)\bar{u}(p_1)=\!\not\!{p}_1+m_1$ and $\sum v(p_2)\bar{v}(p_2)=\!\not\!{p}_2-m_2$ for the particle and antiparticle respectively, and take the notation $K^2=(p_1+p_2)^2$. We carry out  the integrals in Eqs.(17-18) in $D=4+2\varepsilon_{\rm IR}$ dimension as there is only infrared divergence, and obtain the contributions of the real gluon emissions (see Fig.3) to spectral densities,
\begin{eqnarray}
\frac{{\rm Im}\Pi^R_{V/A}(s)}{\pi}&=&\frac{4}{3}\frac{\alpha_s}{\pi}\rho_{V/A}^0(s)\left\{-\frac{1}{\varepsilon_{\rm IR}}+2\log4\pi-2\gamma+\frac{8}{3}-\log\frac{\lambda^3(s,m_1^2,m_2^2)}{m_1^2m_2^2s^2\mu^4} +(s-m_1^2-m_2^2)\overline{R}_{12}(s)\right.\nonumber\\
&&-\overline{R}_{11}(s)-\overline{R}_{22}(s)-\frac{2(s-m_1^2-m_2^2)}{3\sqrt{\lambda(s,m_1^2,m_2^2)}}\log\left(\frac{1+\omega}{1-\omega}\right) -R_{12}^1(s) +\frac{s-m_1^2-m_2^2}{\sqrt{\lambda(s,m_1^2,m_2^2)}} \nonumber\\
&&\left.\log\left(\frac{1+\omega}{1-\omega}\right) \left[\frac{1}{\varepsilon_{\rm IR}}-2\log4\pi+2\gamma-2+\log\frac{\lambda^3(s,m_1^2,m_2^2)}{m_1^2m_2^2s^2\mu^4} \right]\right\}\nonumber\\
&&+\frac{4}{3}\frac{\alpha_s}{\pi}\left\{ \frac{ s-(m_1\mp m_2)^2 }{4s\pi^2}\left[\log\left(\frac{1+\omega}{1-\omega}\right)(s-m_1^2-m_2^2)-\sqrt{\lambda(s,m_1^2,m_2^2)}\right]
    \right.\nonumber\\
&&\left.+\frac{\sqrt{\lambda(s,m_1^2,m_2^2)}}{16s\pi^2} R_{12}^2(s)\left[2+\frac{(m_1\mp m_2)^2}{s}\right]-\frac{1}{\pi^2}R_0(s)\right\} \, ,
\end{eqnarray}
\begin{eqnarray}
\frac{{\rm Im}\Pi^R_{S/P}(s)}{\pi}&=&\frac{4}{3}\frac{\alpha_s}{\pi}\rho_{S/P}^0(s)\left\{-\frac{1}{\varepsilon_{\rm IR}}+2\log4\pi-2\gamma+2-\log\frac{\lambda^3(s,m_1^2,m_2^2)}{m_1^2m_2^2s^2\mu^4} +(s-m_1^2-m_2^2)\overline{R}_{12}(s)\right.\nonumber\\
&&-\overline{R}_{11}(s)-\overline{R}_{22}(s)-R_{12}^1(s) +\frac{R_{12}^2}{2}\frac{1}{s-(m_1\pm m_2)^2}+\frac{s-m_1^2-m_2^2}{\sqrt{\lambda(s,m_1^2,m_2^2)}} \nonumber\\
&&\left.\log\left(\frac{1+\omega}{1-\omega}\right) \left[\frac{1}{\varepsilon_{\rm IR}}-2\log4\pi+2\gamma-2+\log\frac{\lambda^3(s,m_1^2,m_2^2)}{m_1^2m_2^2s^2\mu^4} \right]\right\}\, ,
\end{eqnarray}
the  expressions of the $\overline{R}_{11}(s)$, $\overline{R}_{22}(s)$, $\overline{R}_{12}(s)$, $R_{12}^1(s)$ and $R^2_{12}(s)$ are  given explicitly in the appendix. We have used the $\gamma_5$ symmetry of the interpolating currents to obtain the  spectral densities $\frac{{\rm Im}\Pi^R_{A}(s)}{\pi}$ and $\frac{{\rm Im}\Pi^R_{P}(s)}{\pi}$.

\begin{figure}
 \centering
 \includegraphics[totalheight=3cm,width=7cm]{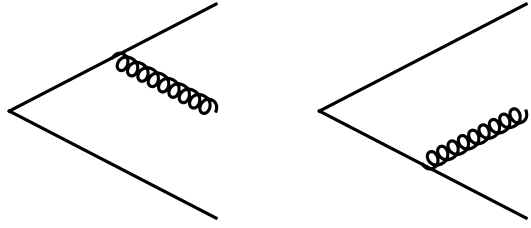}
    \caption{The amplitudes for the real gluon emissions. }
\end{figure}

\section{Analytical expressions of the total contributions}
We add the contributions of  the virtual and real gluon emissions together, and obtain the total perturbative  spectral densities at the next-to-leading order,
 \begin{eqnarray}
\rho_{V/A}^1(s)&=&\frac{4}{3}\frac{\alpha_s}{\pi}\rho_{V/A}^0(s)\left\{ \frac{1}{2}\overline{f}_{\pm}(s)-\overline{R}_{11}(s)-\overline{R}_{22}(s)+(s-m_1^2-m_2^2)\overline{R}_{12}(s)-\frac{5}{6}\right. \nonumber\\
&&+2\log\frac{\sqrt[4]{m_1^7m_2^7s}}{\lambda(s,m_1^2,m_2^2)}+\frac{2(s-m_1^2-m_2^2)}{\sqrt{\lambda(s,m_1^2,m_2^2)}}\log\left(\frac{1+\omega}{1-\omega}\right)
\log\frac{\lambda(s,m_1^2,m_2^2)}{m_1 m_2 s}\nonumber\\
&&\left. -\frac{2(s-m_1^2-m_2^2)}{3\sqrt{\lambda(s,m_1^2,m_2^2)}}\log\left(\frac{1+\omega}{1-\omega}\right)-R_{12}^1(s)\right\} \nonumber\\
&&+\frac{4}{3}\frac{\alpha_s}{\pi}\left\{ \frac{s-(m_1\mp m_2)^2}{4s\pi^2}\left[\log\left(\frac{1+\omega}{1-\omega}\right)(s-m_1^2-m_2^2)-\sqrt{\lambda(s,m_1^2,m_2^2)}\right] \right. \nonumber\\
&&+\frac{\sqrt{\lambda(s,m_1^2,m_2^2)}}{16s\pi^2}R_{12}^2(s)\left[2+\frac{(m_1\mp m_2)^2}{s}\right]-\frac{1}{\pi^2}R_0(s)+\frac{\sqrt{\lambda(s,m_1^2,m_2^2)}^3}{s^2}\nonumber\\
&&\left.\left[\frac{1}{12\pi^2} \left(1-\frac{s-m_1^2-m_2^2}{\sqrt{\lambda(s,m_1^2,m_2^2)}}\log\left(\frac{1+\omega}{1-\omega}\right) \right) -\frac{(m_1\pm m_2)\left(f_{1\pm}(s)+f_{2\pm}(s) \right)}{32\pi^2} \right]\right\} \, ,\nonumber\\
\end{eqnarray}
\begin{eqnarray}
\rho_{S/P}^1(s)&=&\frac{4}{3}\frac{\alpha_s}{\pi}\rho_{S/P}^0(s)\left\{ \frac{1}{2}\overline{f}_{\pm}(s)-\overline{R}_{11}(s)-\overline{R}_{22}(s)+(s-m_1^2-m_2^2)\overline{R}_{12}(s)-\frac{3}{2}\right. \nonumber\\
&&+2\log\frac{\sqrt[4]{m_1^7m_2^7s}}{\lambda(s,m_1^2,m_2^2)}+\frac{2(s-m_1^2-m_2^2)}{\sqrt{\lambda(s,m_1^2,m_2^2)}}\log\left(\frac{1+\omega}{1-\omega}\right)
\log\frac{\lambda(s,m_1^2,m_2^2)}{m_1 m_2 s}\nonumber\\
&&\left. -R_{12}^1(s)+\frac{R_{12}^2(s)}{2}\frac{1}{s-(m_1\pm m_2)^2}\right\} \nonumber\\
&&+\frac{4}{3}\frac{\alpha_s}{\pi}\frac{3}{32\pi^2}\frac{\sqrt{\lambda(s,m_1^2,m_2^2)}}{s^2}\left\{(s+m_1^2-m_2^2)(s+m_2^2-m_1^2)\left[f_{1\pm}(s)m_1\pm f_{2\pm}(s)m_2\right]  \right.\nonumber\\
&&\left.\mp f_{1\pm}(s)m_2(s+m_1^2-m_2^2)^2-f_{2\pm}(s)m_1(s+m_2^2-m_1^2)^2 \right\} \, .
\end{eqnarray}
The infrared divergences $\frac{1}{\varepsilon_{\rm IR}}$, $\log\left(\frac{1+\omega}{1-\omega} \right)\frac{1}{\varepsilon_{\rm IR}}$   from the virtual and real gluon emissions  are canceled out with each other, which is guaranteed by the Lee-Nauenberg theorem \cite{LeeBook}. The spectral densities at the next-to-leading order do not have renormalization scale dependence, although the strong coupling constant $\alpha_s(\mu)$ is energy scale dependent quantity.  In the appendix, we present some necessary technical details for readers convenience, one can follow the necessary steps to check the calculations if one does not feel confident with the final expressions.

We compare the spectral densities of the order ${\mathcal{O}}(\alpha_s)$  numerically to other results in Refs.\cite{Djouadi1994,Groote1301} by taking the energy scale $\mu=5\,\rm{GeV}$, pole masses $m_1=m_b=4.8\,\rm{GeV}$, $m_2=m_c=1.5\,\rm{GeV}$, and  observe that there are differences that cannot be neglected. We can use the present expressions  to study the masses and decay constants of the scalar, pseudoscalar, vector and axialvector $B_c$ mesons with the QCD sum rules in a systematic way.

\section{Conclusion}
In this article, we calculate the next-to-leading order perturbative contributions  to the spectral densities of the mesonic two-point correlation functions of the vector, axialvector, scalar and pseudoscalar currents. In calculations, we assume that the quarks have unequal masses and  use optical theorem to obtain the spectral densities directly, furthermore, we take the scheme of the dimensional regularization to regularize both the ultraviolet and infrared divergences. The ultraviolet and infrared divergent terms are canceled out with each other separately, the net spectral densities are free of divergences. The analytical expressions are applicable in  many phenomenological analysis besides  the QCD sum rules.

\section*{Acknowledgements}
This  work is supported by National Natural Science Foundation,
Grant Numbers 11075053, 11375063,  and the Fundamental Research Funds for the
Central Universities.

\section*{Appendix}
Firstly, we write down the fundamental integrals in calculating the vertex corrections,
\begin{eqnarray}
V_{ab}(s)&=& 16\pi^2\int_0^1 dx \int_0^{1-x}dy \int \frac{d^D k_E}{(2\pi)^D} \frac{x^a y^b \Gamma(3)}{\left[k^2_E+(x p_1+y p_2)^2\right]^3}  \, ,\nonumber\\
V(s)&=& 16\pi^2\left(1-\frac{3}{2}\varepsilon_{\rm UV}\right)\int_0^1 dx \int_0^{1-x}dy \int \frac{d^D k_E}{(2\pi)^D}
\frac{k^2_E \Gamma(3)}{\left[k^2_E+(x p_1+y p_2)^2\right]^3} \, ,
\end{eqnarray}
and carry out the integrals to obtain the following analytical expressions,
\begin{eqnarray}
V_{00}(s)&=& \frac{1}{\sqrt{\lambda(s,m_1^2,m_2^2)}}\left\{ -\log\left(\frac{1+\omega}{1-\omega}\right)\left(\frac{1}{\varepsilon_{\rm IR}}+\log\frac{s}{4\pi\mu^2}+\gamma \right) + \frac{\log^2(1-\omega_1^2)}{4}-\log^2(1+\omega_1) \right. \nonumber \\
&&+ \frac{\log^2(1-\omega_2^2)}{4}-\log^2(1+\omega_2) +2\log(\omega_1+\omega_2)\log\left(\frac{1+\omega}{1-\omega}\right)-\log\omega_1\log\left(\frac{1+\omega_2}{1-\omega_2}\right) \nonumber \\
&&\left.-\log\omega_2\log\left(\frac{1+\omega_1}{1-\omega_1}\right)-{\rm Li_2}\left( \frac{2\omega_1}{1+\omega_1} \right) -{\rm Li_2}\left( \frac{2\omega_2}{1+\omega_2} \right)+\pi^2\right\} \, ,\nonumber \\
&=&\overline{V}_{00}(s)- \frac{1}{\sqrt{\lambda(s,m_1^2,m_2^2)}}\log\left(\frac{1+\omega}{1-\omega}\right)\left(\frac{1}{\varepsilon_{\rm IR}}+\log\frac{s}{4\pi\mu^2}+\gamma \right) \, , \nonumber
\end{eqnarray}
\begin{eqnarray}
V_{10}(s)&=& \frac{1}{s}\left\{\frac{1}{2}\log\left(\frac{1-\omega_1^2}{1-\omega_2^2}\right)-\frac{1}{\omega_2}\log\left(\frac{1+\omega}{1-\omega} \right)+\log\frac{\omega_2}{\omega_1}\right\} \, ,\nonumber \\
V_{01}(s)&=& V_{10}(s)|_{\omega_1 \leftrightarrow \omega_2} \, ,\nonumber \\
V_{20}(s)&=& \frac{1}{2s}\left\{-\frac{\omega_1\omega_2}{\omega_1+\omega_2}\log\left(\frac{1+\omega}{1-\omega} \right)-\frac{\omega_1}{\omega_2(\omega_1+\omega_2)}\log\left(\frac{1+\omega}{1-\omega} \right)+   \frac{\omega_1}{\omega_1+\omega_2}\log\left(\frac{1-\omega_1^2}{1-\omega_2^2}\right)\right.\nonumber\\
&&\left.+\frac{2\omega_1}{\omega_1+\omega_2}\log\frac{\omega_2}{\omega_1}+1\right\} \, ,\nonumber \\
V_{02}(s)&=&V_{20}(s)|_{\omega_1 \leftrightarrow \omega_2} \, ,\nonumber \\
V_{11}(s)&=& \frac{1}{2s}\left\{\frac{\omega_1\omega_2}{\omega_1+\omega_2}\log\left(\frac{1+\omega}{1-\omega} \right)-  \frac{\omega_1-\omega_2}{2(\omega_1+\omega_2)}\log\left(\frac{1-\omega_1^2}{1-\omega_2^2}\right)-\frac{1}{\omega_1+\omega_2}\log\left(\frac{1+\omega}{1-\omega} \right)\right.\nonumber\\
&&\left.+\frac{\omega_1}{\omega_1+\omega_2}\log\frac{\omega_1}{\omega_2}+\frac{\omega_2}{\omega_1+\omega_2}\log\frac{\omega_2}{\omega_1}-1\right\} \, ,\nonumber
\end{eqnarray}
\begin{eqnarray}
V(s)&=& \frac{1}{\varepsilon_{\rm UV}}+\log\frac{4\pi\mu^2}{s}-\gamma+1-\frac{2\omega_1\omega_2}{\omega_1+\omega_2}\log\left(\frac{1+\omega}{1-\omega} \right)
-\frac{\omega_2}{\omega_1+\omega_2}\log(1-\omega_1^2)\nonumber \\
&&-\frac{\omega_1}{\omega_1+\omega_2}\log(1-\omega_2^2)-2\frac{\omega_1\log\omega_1+\omega_2\log\omega_2}{\omega_1+\omega_2}+2\log(\omega_1+\omega_2)\, , \nonumber\\
&=& \overline{V}(s)+ \frac{1}{\varepsilon_{\rm UV}}+\log\frac{4\pi\mu^2}{s}-\gamma+1\, ,
\end{eqnarray}
where
\begin{eqnarray}
\omega_1&=&\frac{\sqrt{\lambda(s,m_1^2,m_2^2)}}{s+m_1^2-m_2^2} \, ,\nonumber\\
\omega_2&=&\frac{\sqrt{\lambda(s,m_1^2,m_2^2)}}{s+m_2^2-m_1^2} \, ,\nonumber\\
M&=&\frac{m_1+m_2}{m_1-m_2}\, ,\nonumber\\
{\rm Li_2}(x)&=&-\int_0^x dt \frac{\log(1-t)}{t} \, .
\end{eqnarray}

Secondly, we take the notation
\begin{eqnarray}
\int d ps&=& \int \frac{d^{D-1}\vec{k}}{2E_k}\frac{d^{D-1}\vec{p}_1}{2E_{p_1}}\frac{d^{D-1}\vec{p}_2}{2E_{p_2}}\delta^D(p-k-p_1-p_2) \, ,\nonumber
\end{eqnarray}
for simplicity, and write down the analytical expressions of the three-body phase-space integrals,
  \begin{eqnarray}
R_0(s)&=& \frac{1}{\pi^2}\int d ps\nonumber\\
&=&\frac{m_1^2-m_2^2}{4}\log\left(\frac{M+\omega}{M-\omega}\right)-\frac{s m_1^2+s m_2^2-2m_1^2m_2^2}{4s}\log\left(\frac{1+\omega}{1-\omega}\right)\nonumber\\
&&+\frac{\sqrt{\lambda(s,m_1^2,m_2^2)}(s+m_1^2+m_2^2)}{8s}\, , \nonumber
\end{eqnarray}
\begin{eqnarray}
 R_{11}(s)&=& \frac{s m_1^2}{\pi^2\sqrt{\lambda(s,m_1^2,m_2^2)}}(2\pi)^{-4\varepsilon_{\rm IR}}\mu^{-2\varepsilon_{\rm IR}}\int d ps \frac{1}{(k \cdot p_1 )^2}\nonumber\\
&=& \frac{1}{2\varepsilon_{\rm IR}}-\log4\pi+\gamma-1+\log\frac{\sqrt{\lambda(s,m_1^2,m_2^2)}^3}{m_1m_2s\mu^2}-\frac{s+m_1^2-m_2^2}{2\sqrt{\lambda(s,m_1^2,m_2^2)}} \log\left(\frac{1+\omega_1}{1-\omega_1}\right) \nonumber \\
&&-\frac{m_1^2-m_2^2}{\sqrt{\lambda(s,m_1^2,m_2^2)}} \log\left(\frac{1+\omega_1}{1-\omega_1}\right) -\frac{s-m_1^2+m_2^2}{\sqrt{\lambda(s,m_1^2,m_2^2)}} \log\left(\frac{1+\omega}{1-\omega}\right) \nonumber \\
&=&\overline{ R}_{11}(s)+\frac{1}{2\varepsilon_{\rm IR}}-\log4\pi+\gamma-1+\log\frac{\sqrt{\lambda(s,m_1^2,m_2^2)}^3}{m_1m_2s\mu^2} \, , \nonumber \\
R_{22}(s)&=&R_{11}(s)|_{m_1\leftrightarrow m_2} \, , \nonumber
\end{eqnarray}
\begin{eqnarray}
R_{12}(s)&=& \frac{s  }{\pi^2\sqrt{\lambda(s,m_1^2,m_2^2)}}(2\pi)^{-4\varepsilon_{\rm IR}}\mu^{-2\varepsilon_{\rm IR}}\int d ps \frac{1}{k\cdot p_1 k\cdot p_2 }\nonumber\\
&=&\frac{1}{\sqrt{\lambda(s,m_1^2,m_2^2)}} \left\{\log\left(\frac{1+\omega}{1-\omega}\right)\left[ \frac{1}{\varepsilon_{\rm IR}}-2\log4\pi+2\gamma-2+2\log\frac{\sqrt{\lambda(s,m_1^2,m_2^2)}^3}{m_1m_2s\mu^2}\right] \right. \nonumber \\
&&-2\log\frac{m_1}{m_2}\log\left(\frac{M+\omega}{M-\omega}\right)-\log^2\left(\frac{1+\omega}{1-\omega}\right)+2\log\frac{s}{\bar{s}}\log\left(\frac{1+\omega}{1-\omega}\right)
-4{\rm Li_2}\left( \frac{2\omega}{1+\omega}\right) \nonumber\\
&&+2{\rm Li_2}\left( \frac{\omega-1}{\omega-M}\right) +2{\rm Li_2}\left( \frac{\omega-1}{\omega+M}\right)-2{\rm Li_2}\left( \frac{\omega+1}{\omega-M}\right)-2{\rm Li_2}\left( \frac{\omega+1}{\omega+M}\right)-\frac{1}{2}{\rm Li_2}\left( \frac{1+\omega_1}{2}\right)\nonumber\\
&&\left.-\frac{1}{2}{\rm Li_2}\left( \frac{1+\omega_2}{2}\right)-{\rm Li_2}\left( \omega_1\right)-{\rm Li_2}\left( \omega_2\right)+\frac{\log2\log\left[(1+\omega_1)(1+\omega_2) \right]}{2}-\frac{\log^2 2}{2}+\frac{\pi^2}{12}\right\}\, ,\nonumber\\
&=&\overline{R}_{12}(s)+\frac{1}{\sqrt{\lambda(s,m_1^2,m_2^2)}} \log\left(\frac{1+\omega}{1-\omega}\right)\left[ \frac{1}{\varepsilon_{\rm IR}}-2\log4\pi+2\gamma-2+2\log\frac{\sqrt{\lambda(s,m_1^2,m_2^2)}^3}{m_1m_2s\mu^2}\right] \, , \nonumber
\end{eqnarray}
\begin{eqnarray}
R^1_{12}(s)&=& \frac{s  }{\pi^2\sqrt{\lambda(s,m_1^2,m_2^2)}}\int d ps \frac{s-K^2}{k\cdot p_1 k\cdot p_2 }\nonumber\\
&=&\frac{s}{\sqrt{\lambda(s,m_1^2,m_2^2)}} \left\{\log^2(1-\omega)-\log^2(1+\omega) +2\log\frac{2s}{\bar{s}}\log\left(\frac{1+\omega}{1-\omega}\right)
+2{\rm Li_2}\left(\frac{1-\omega}{2}\right)\right. \nonumber \\
&&\left.-2{\rm Li_2}\left(\frac{1+\omega}{2}\right)+2{\rm Li_2}\left(\frac{1+\omega}{1+M}\right)+2{\rm Li_2}\left(\frac{1+\omega}{1-M}\right)
-2{\rm Li_2}\left(\frac{1-\omega}{1-M}\right)-2{\rm Li_2}\left(\frac{1-\omega}{1+M}\right)\right\} \, ,\nonumber \\
R^2_{12}(s)&=& \frac{s}{\pi^2\sqrt{\lambda(s,m_1^2,m_2^2)}}\int d ps \frac{\left(s-K^2\right)^2}{k\cdot p_1 k\cdot p_2 }\nonumber\\
&=&\frac{s^2}{\sqrt{\lambda(s,m_1^2,m_2^2)}} \left\{\log^2(1-\omega)-\log^2(1+\omega) +2\log\frac{4s}{\bar{s}}\log\left(\frac{1+\omega}{1-\omega}\right)
+2{\rm Li_2}\left(\frac{1-\omega}{2}\right)\right. \nonumber \\
&&-2{\rm Li_2}\left(\frac{1+\omega}{2}\right)+2{\rm Li_2}\left(\frac{1+\omega}{1+M}\right)+2{\rm Li_2}\left(\frac{1+\omega}{1-M}\right)
-2{\rm Li_2}\left(\frac{1-\omega}{1-M}\right)-2{\rm Li_2}\left(\frac{1-\omega}{1+M}\right) \, \nonumber \\
&&\left. +\frac{2\omega \bar{s}}{s} -\frac{\bar{s}}{s}(1+\omega^2)\log\left(\frac{1+\omega}{1-\omega}\right) \right\} \, ,
\end{eqnarray}
where $\bar{s}=s-(m_1-m_2)^2$.

In the following, we present some necessary technical details in calculating the integrals for both the virtual and real gluon emissions, one can check the calculations; here we smear the lower index $\rm {IR}$ in $ \varepsilon_{\rm IR}$ for simplicity.
\begin{eqnarray}
\int_0^1 dx \int_0^{1-x}dy \int \frac{d^D k_E}{(2\pi)^D} \frac{ \Gamma(3)}{\left[k^2_E+(x p_1+y p_2)^2\right]^3}&=&\int_0^1 dx \int_0^{1-x}dy   \frac{ (4\pi)^{-\varepsilon}\Gamma(1-\varepsilon)}{ 16\pi^2(x p_1+y p_2)^{2-2\varepsilon} } \nonumber
\end{eqnarray}
\begin{eqnarray}
&=&\int_0^1 dz \int_0^1 dtt \frac{(4\pi)^{-\varepsilon}\Gamma(1-\varepsilon)}{16\pi^2\left\{t^2 s \left[z-\frac{\omega_1(1+\omega_2)}{\omega_1+\omega_2} \right]\left[z-\frac{\omega_1(1-\omega_2)}{\omega_1+\omega_2} \right]\right\}^{1-\varepsilon}} \nonumber\\
&=& \frac{1}{16\pi^2\sqrt{\lambda(s,m_1^2,m_2^2)}}\left\{ -\log\left(\frac{1+\omega}{1-\omega}\right)\left(\frac{1}{\varepsilon}+\log\frac{s}{4\pi\mu^2}+\gamma \right) + \frac{\log^2(1-\omega_1^2)}{4}-\log^2(1+\omega_1) \right. \nonumber \\
&&+ \frac{\log^2(1-\omega_2^2)}{4}-\log^2(1+\omega_2) +2\log(\omega_1+\omega_2)\log\left(\frac{1+\omega}{1-\omega}\right)-\log\omega_1\log\left(\frac{1+\omega_2}{1-\omega_2}\right) \nonumber \\
&&\left.-\log\omega_2\log\left(\frac{1+\omega_1}{1-\omega_1}\right)-{\rm Li_2}\left( \frac{2\omega_1}{1+\omega_1} \right) -{\rm Li_2}\left( \frac{2\omega_2}{1+\omega_2} \right)+\pi^2\right\} \, .
\end{eqnarray}

\begin{eqnarray}
\int d ps \frac{\mu^{-2\varepsilon}}{k\cdot p_1 k\cdot p_2 }&=&\frac{\pi^{\frac{5}{2}+2\varepsilon}\mu^{-2\varepsilon}}{\Gamma(1+\varepsilon)\Gamma(\frac{3}{2}+\varepsilon)}\int_{(m_1+m_2)^2}^s dK^2 \frac{(s-K^2)^{1+2\varepsilon}}{(4s)^{1+\varepsilon}} \frac{\sqrt{\lambda(K^2,m_1^2,m_2^2)}^{1+2\varepsilon}}{(4K^2)^{1+\varepsilon}} \int_{-1}^1 dcos\theta \nonumber\\
&& \frac{(1-cos^2\theta)^\varepsilon}{\left[\frac{s-K^2}{2\sqrt{s}}\right]^2\left[\frac{K^2+m_1^2-m_2^2}{2K}-\frac{\sqrt{\lambda(K^2,m_1^2,m_2^2)}}{2K}cos\theta\right]
\left[\frac{K^2+m_2^2-m_1^2}{2K}+\frac{\sqrt{\lambda(K^2,m_1^2,m_2^2)}}{2K}cos\theta\right]}\nonumber
\end{eqnarray}
\begin{eqnarray}
&=&\frac{\pi^{3+2\varepsilon}\mu^{-2\varepsilon}}{\Gamma(\frac{3}{2}+\varepsilon)\Gamma(\frac{3}{2}+\varepsilon)}\int_{(m_1+m_2)^2}^s dK^2 \frac{1}{4K^2(s-K^2)^{1-2\varepsilon}}  \left[1+\varepsilon\log\frac{\lambda(K^2,m_1^2,m_2^2)}{16sK^2} \right] \nonumber\\
&& \left\{ 2\log\left(\frac{1+u}{1-u}\right)-\varepsilon \left[ {\rm Li_2}\left(\frac{1+u_1}{2}\right)+{\rm Li_2}\left(\frac{1+u_2}{2}\right) +2{\rm Li_2} (u_1)+2{\rm Li_2}(u_2) \right.\right.\nonumber\\
&&\left.\left.-4\log\left(\frac{1+u}{1-u} \right)-\log2\log\left[ (1+u_1)(1+u_2)\right]+\log^{2}2-\frac{\pi^2}{6} \right]\right\}\nonumber\\
&=&\frac{\pi^{3+2\varepsilon}\mu^{-2\varepsilon}}{\Gamma(\frac{3}{2}+\varepsilon)\Gamma(\frac{3}{2}+\varepsilon)}\frac{2m_1m_2}{(m_1-m_2)^2\left[s-(m_1-m_2)^2\right]^{1-2\varepsilon}}
\int_{0}^\omega du \frac{u}{(M^2-u^2)(\omega^2-u^2)^{1-2\varepsilon}}  \nonumber\\
&& \left[1+\varepsilon\log\frac{\lambda(K^2,m_1^2,m_2^2)}{16sK^2} -2\varepsilon\log(1-u^2)\right] \left\{ 2\int_{-1}^1 dx \frac{u}{u^2-x^2}-\varepsilon \left[ {\rm Li_2}\left(\frac{1+u_1}{2}\right) +{\rm Li_2}\left(\frac{1+u_2}{2}\right)\right.\right.\nonumber\\
&&\left.\left. +2{\rm Li_2} (u_1)+2{\rm Li_2}(u_2)-4\log\left(\frac{1+u}{1-u} \right)
-\log2\log\left[ (1+u_1)(1+u_2)\right]+\log^{2}2-\frac{\pi^2}{6} \right]\right\}\nonumber
\end{eqnarray}
\begin{eqnarray}
&=&\frac{\pi^2}{s} \left\{\log\left(\frac{1+\omega}{1-\omega}\right)\left[ \frac{1}{\varepsilon}+2\log\pi+2\gamma-2+2\log\frac{\sqrt{\lambda(s,m_1^2,m_2^2)}^3}{m_1m_2s\mu^2}\right]-2\log\frac{m_1}{m_2}\log\left(\frac{M+\omega}{M-\omega}\right) \right. \nonumber \\
&&-\log^2\left(\frac{1+\omega}{1-\omega}\right)+2\log\frac{s}{\bar{s}}\log\left(\frac{1+\omega}{1-\omega}\right)
-4{\rm Li_2}\left( \frac{2\omega}{1+\omega}\right)+2{\rm Li_2}\left( \frac{\omega-1}{\omega-M}\right) +2{\rm Li_2}\left( \frac{\omega-1}{\omega+M}\right) \nonumber\\
&&-2{\rm Li_2}\left( \frac{\omega+1}{\omega-M}\right)-2{\rm Li_2}\left( \frac{\omega+1}{\omega+M}\right)-\frac{1}{2}{\rm Li_2}\left( \frac{1+\omega_1}{2}\right)-\frac{1}{2}{\rm Li_2}\left( \frac{1+\omega_2}{2}\right)-{\rm Li_2}\left( \omega_1\right)-{\rm Li_2}\left( \omega_2\right)\nonumber\\
&&\left.+\frac{\log2\log\left[(1+\omega_1)(1+\omega_2) \right]}{2}-\frac{\log^2 2}{2}+\frac{\pi^2}{12}\right\} \,  ,
\end{eqnarray}
where
\begin{eqnarray}
u_1&=&\frac{\sqrt{\lambda(K^2,m_1^2,m_2^2)}}{K^2+m_1^2-m_2^2} \, ,\nonumber\\
u_2&=&\frac{\sqrt{\lambda(K^2,m_1^2,m_2^2)}}{K^2+m_2^2-m_1^2} \, ,\nonumber\\
u&=&\sqrt{\frac{K^2-(m_1+m_2)^2}{K^2-(m_1-m_2)^2}}\, ,
\end{eqnarray}
and we have neglected the imaginary terms.
The infrared divergent terms of the form $\log\left(\frac{1+\omega}{1-\omega}\right)\frac{1}{\varepsilon}$ come from  the diagrams of the
 virtual gluon emissions (vertex corrections) and real gluon emissions are canceled out with each other.

In calculating the divergent integrals, we have used the following trick,
\begin{eqnarray}
\int_0^\omega dx \frac{f(x,\varepsilon)}{(\omega-x)^{1-\varepsilon}}\left[1+\varepsilon g(x)\right]&=&\int_0^\omega dx \frac{f(x,\varepsilon)}{(\omega-x)^{1-\varepsilon}}+\varepsilon\int_0^\omega dx \frac{f(x,\varepsilon)\left[g(x)-g(\omega)\right]}{(\omega-x)^{1-\varepsilon}}+\varepsilon\int_0^\omega dx \frac{f(x,\varepsilon)g(\omega)}{(\omega-x)^{1-\varepsilon}}\nonumber\\
&=&\int_0^\omega dx \frac{f(x,\varepsilon)}{(\omega-x)^{1-\varepsilon}}+\varepsilon\int_0^\omega dx \frac{f(x)\left[g(x)-g(\omega)\right]}{\omega-x}+\varepsilon\int_0^\omega dx \frac{f(x,\varepsilon)g(\omega)}{(\omega-x)^{1-\varepsilon}}\nonumber\\
&=&\int_0^\omega dx \frac{f(x,\varepsilon)}{(\omega-x)^{1-\varepsilon}}+\varepsilon\int_0^\omega dx \frac{f(x,\varepsilon)g(\omega)}{(\omega-x)^{1-\varepsilon}}\, ,
\end{eqnarray}
the functions $f(x,\varepsilon)$ and $g(x)$ have no poles in the range $x=0-\omega$ and $\varepsilon\rightarrow 0^+$.

\end{document}